\documentclass[useAMS,usenatbib]{mn2e}

\usepackage{graphicx,amssymb,color}
\usepackage[normalem]{ulem}

\title[Central configurations and planetary systems]
{Relating binary-star planetary systems to central configurations}

\author[Veras]{
Dimitri Veras$^{1}$\thanks{E-mail: d.veras@warwick.ac.uk}
\\
$^{1}$Department of Physics, University of Warwick, Coventry CV4 7AL, UK
}

\pubyear{2016}

\begin{document}
\label{firstpage}
\pagerange{\pageref{firstpage}--\pageref{lastpage}}
\maketitle

\begin{abstract}
Binary-star exoplanetary systems are now known to be common, for both wide and close binaries. However, their orbital evolution is generally unsolvable. Special cases of the $N$-body problem which are in fact completely solvable include dynamical architectures known as central configurations. Here, I utilize recent advances in our knowledge of central configurations to assess the plausibility of linking them to coplanar exoplanetary binary systems. By simply restricting constituent masses to be within stellar or substellar ranges characteristic of planetary systems, I find that (i) this constraint reduces by over 90 per cent the phase space in which central configurations may occur, (ii) both equal-mass and unequal-mass binary stars admit central configurations, (iii) these configurations effectively represent different geometrical extensions of the Sun-Jupiter-Trojan-like architecture, (iv) deviations from these geometries are no greater than ten degrees, and (v) the deviation increases as the substellar masses increase. This study may help restrict future stability analyses to architectures which resemble exoplanetary systems, and might hint at where observers may discover dust, asteroids and/or planets in binary star systems.
\end{abstract}

\begin{keywords}
  celestial mechanics
  --
  stars: binaries: general
  --
  minor planets, asteroids: general
  --
  stars: kinematics and dynamics
  --
  planet and satellites: dynamical evolution and stability
  --
  protoplanetary discs
\end{keywords}

\section{Introduction}

The Trojan asteroids hosted by Jupiter and Neptune help constrain the formation of the Solar System \citep{chilit2005,morbidelli2005,lyketal2009} and represent powerful examples of how central configurations -- a topic often limited to the mathematics literature -- apply to a real planetary system. Extrasolar planetary systems potentially provide other opportunities, with their diverse architectures, and our increasing capacity to observe sub-Earth mass solid bodies \citep[e.g.][]{kieetal2014,vanetal2015} and clumps of dust \citep[e.g.][]{rapetal2015,haretal2016}. 

Central configurations produce exact solutions to the $N$-body problem, which is otherwise generally unsolvable. Hence, characterizing the existence and architectures of all central configurations for all $N$ is a holy grail of celestial mechanics (see \citealt*{moeckel1990} and Chapter 2 of \citealt*{libetal2015}). They can aid in understanding the long-term dynamics of planetary systems, identifying the masses and locations of objects which remain stable, and targeting searches for hidden objects. Strictly, a central configuration is an arrangement of point masses which satisfy

\begin{equation}
\Lambda \vec{r}_i = \sum_{j \ne i}^{N} \frac{M_j \left(\vec{r}_j - \vec{r}_i \right)}{r_{ij}^3}
,
\label{orig}
\end{equation}

\noindent{}where $\vec{r}$ is the barycentric position vector, $M$ is mass, $i,j$ are object indices, $r_{ij} = |\vec{r}_i - \vec{r}_j|$, $N$ is the total number of bodies, and $\Lambda$ is a constant.

When $N=2$, all systems are central configurations. For $N=3$, there exists two classes of central configurations: when all three bodies are co-linear \citep{euler1767} and when all three bodies form an equilateral triangle \citep{lagrange1772}. Only some co-linear systems (for correctly chosen masses and distances) are central configurations, whereas all equilateral triangle configurations (for any masses or distances) are central configurations. Co-linear central configurations give rise to the quintic equation which yields Hill stability boundaries \citep{marboz1982}, a now well-used concept in exoplanetary dynamics \citep{davetal2014}. A stability analysis of equilateral triangle architectures reveals the beautiful result that because the Jupiter/Sun mass ratio is less than $(1/2 - \sqrt{23/108}) \approx 0.039$, Jupiter's Trojan asteroids can remain in stable orbits.

\begin{figure*}
\includegraphics[width=18cm]{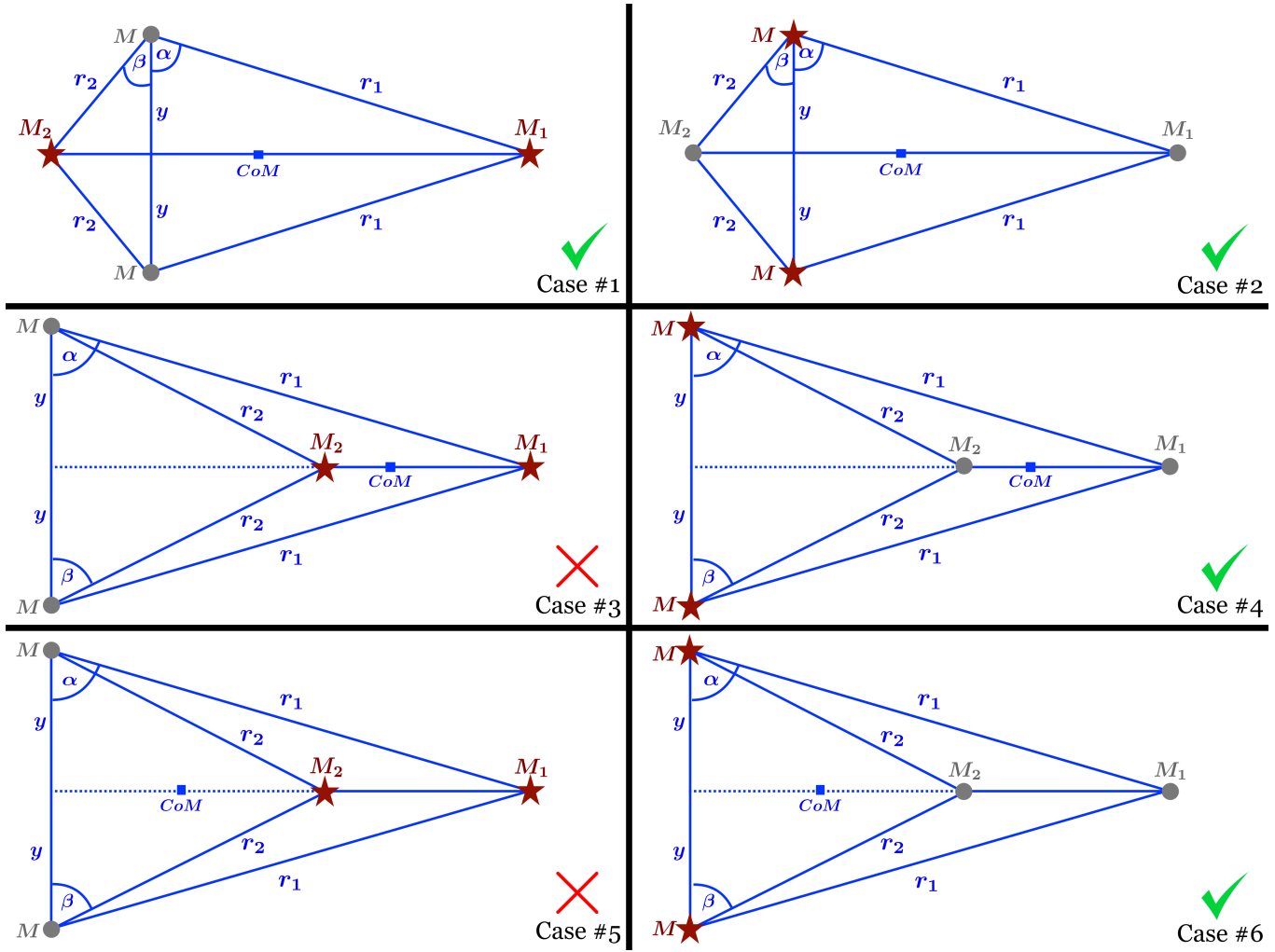}
\caption{Six cases of planar four-body architectures containing two stars
and one axis of symmetry. Stars are labelled in red and with five-pointed
star symbols, and have either unequal masses (left-hand panels) or equal masses (right-hand
panels). The green check marks indicate which cases were found to contain central configurations that 
are potentially applicable to binary star planetary systems.
}
\label{sumfig}
\end{figure*}

\begin{figure*}
\centerline{
\includegraphics[width=8cm]{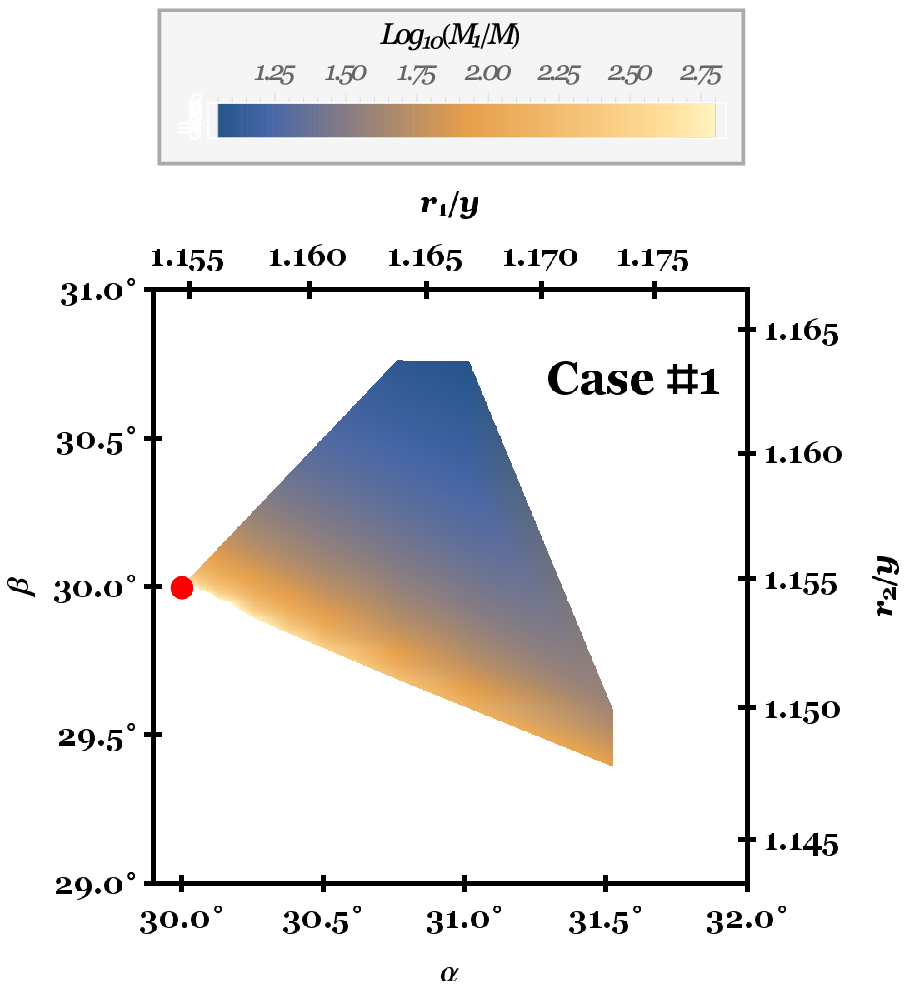}
\ \ \
\includegraphics[width=8cm]{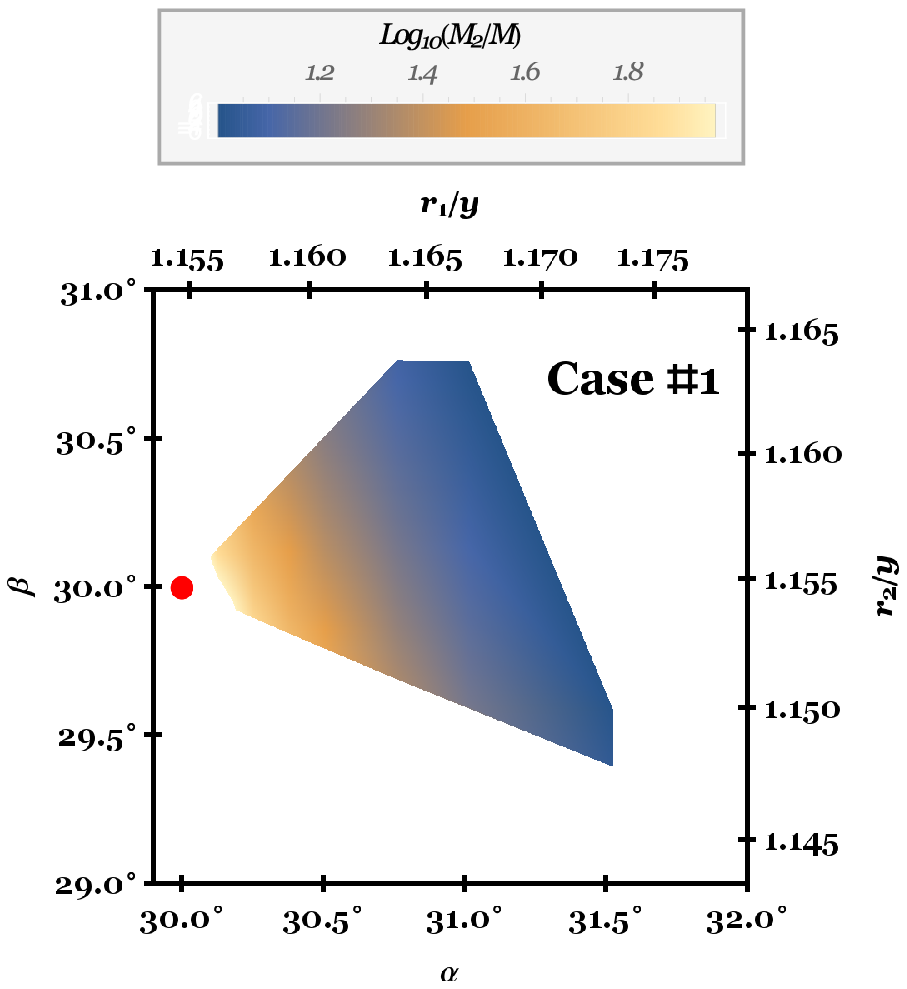}
}
\caption{
The allowed Case \#1 regions that can produce central configurations in binary-star
planetary systems. The plots demonstrate that $\alpha$ and $\beta$ are each limited to
within a few degrees of $30^{\circ}$. The colour contours indicate how much more
restrictive this range is for different object masses: For $M_2/M > 10^2$, both
$\alpha$ and $\beta$ are confined to within a tenth of a degree of $30^{\circ}$,
whereas for $M_1/M > 10^2$, $\alpha$ and $\beta$ lie along the lower axis of the
allowed region.  Asteroids, pebbles or dust (with $M_1/M \gg 10^{10}, M_2/M \gg 10^{10}$) 
effectively lie at the critical point indicated by the red dot.
}
\label{case1mass}
\end{figure*}

\begin{figure}
\includegraphics[width=8cm]{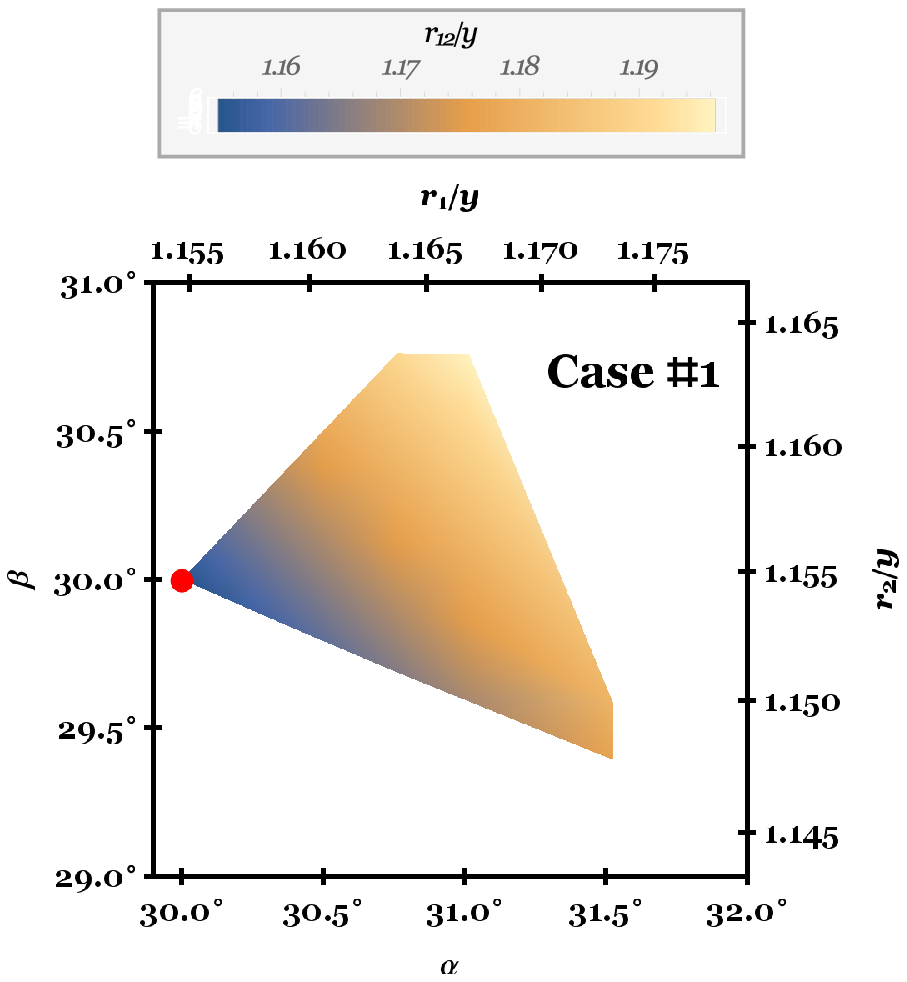}
\caption{
The distance between both stars in Case \#1. Throughout the region
which allows for central configurations to occur, both stars
form a near-equilateral triangle shape (with side-length ratios differing
by no more than five per cent) with each of the other objects.
}
\label{case1distance}
\end{figure}

The $N=4$ case is considerably more complex, but may lend itself well to binary stellar systems which contain dust, minor planets or major planets. A recent significant advancement in the $N=4$ case was provided by \cite{erdczi2016}, who fully characterized all central configurations of four planar bodies, at least two of which contain equal masses, and with an axis of symmetry between the other two bodies. The importance of the result was promoted by \cite{hamilton2016}, who suggested several potential mathematical extensions.

Rather than pursue any of those perhaps daunting challenges, I simply wish here to take stock of the results of \cite{erdczi2016}, and determine broadly if and how they may be applicable to exoplanetary systems, in particular those containing two stars. I cover all four-body planar cases with one axis of symmetry and two equal masses about that axis. The two stars are not restricted to lie along this axis, nor restricted to have equal masses. 

First, in Section 2, I repackage the method for determining these central configurations and provide a straightforward algorithm for the computation. Then, in Section 3 I restrict this class of configurations to masses and distances that correspond to realistic astronomical systems. Doing so allows me to pinpoint, robustly, architectures that might warrant future studies by exoplanet theorists and observers alike. I discuss the implications of this study in Section 4, and then briefly conclude in Section 5.

\section{Computing $N=4$ central configurations}

Consider four objects with masses $M_1$, $M_2$, $M$ and $M$, of which exactly
two are stars. Assume that $M_1 \ne M_2$, $M_1$~$\ne$~$M$ and $M_2 \ne M$. Consequently,
the two stars either have unequal masses $M_1$ and $M_2$ or equal masses $M$. For now,
let the masses of the other two objects be arbitrary. Further, assume a line
of symmetry passes through $M_1$ and $M_2$.

These assumptions yield the six cases shown in Fig. \ref{sumfig}. The stars are indicated
by red five-pointed stars, and the two other objects are gray dots. In the left-hand panels,
the axis of symmetry connects both stars, whereas in the right-hand panels, this axis
connects the other two objects. The difference between the middle and lower panels is
the location of the centre-of-mass (``CoM'') of the systems. In all cases, the distance
between either $M$ object and $M_1$ is $r_1$, and the distance
between either $M$ object and $M_2$ is $r_2$. In the left-hand panels, the distance between
the $M$ objects is $2y$, whereas in the right-hand panels the distance between both stars
is $2y$. \cite{erdczi2016} designated Cases \#1 and \#2 as ``convex'', Cases \#3 and \#4 as
``the first concave case'' and Cases \#5 and \#6 as ``the second concave case''.

As demonstrated by \cite{erdczi2016}, if a configuration is central, then one 
may obtain explicit analytical
expressions for the mass ratios in terms of the geometrical angles $\alpha$ 
and $\beta$ alone. These expressions differ by case, and in each instance
there are restrictions on $\alpha$ and $\beta$.

\subsection{Cases \#1-\#2}

\subsubsection{Masses}

I have found that one
may obtain relatively compact explicit algebraic relations for central
configurations by combining equations 
36, 54-57, and 88 of \cite{erdczi2016}. Doing so gives, for Cases \#1 and \#2,

\begin{eqnarray}
\frac{M_1}{M} &=& \frac{
\tan{\beta} \left(\tan{\alpha} + \tan{\beta}\right)^2 \left(8 \cos^3{\beta} - 1 \right)
}
{
4\left[ \left(\sin{\alpha} + \cos{\alpha} \tan{\beta} \right)^3 -1\right]
}
,
\label{M1abCase12}
\\
\frac{M_2}{M} &=& \frac{
\tan{\alpha} \left(\tan{\alpha} + \tan{\beta}\right)^2 \left(8 \cos^3{\alpha} - 1 \right)
}
{
4\left[ \left(\sin{\beta} + \cos{\beta} \tan{\alpha} \right)^3 - 1\right]
}
,
\label{M2abCase12}
\end{eqnarray}

\noindent{}subject to the restrictions 

\begin{eqnarray}
\alpha &=& 30^{\circ} + 2 \kappa
,
\label{alpha12}
\\
\beta &=& 30^{\circ} + \lambda \kappa
,
\label{beta12}
\end{eqnarray}

\noindent{where}

\begin{equation}
\ \ \, 0 \le \kappa \le 15^{\circ}
,
\label{kappa12}
\end{equation}
\begin{equation}
-1 \le \lambda \le 2
.
\label{lambda12}
\end{equation}

\subsubsection{Distances}
The relative distances between bodies can be expressed in terms
of $\alpha$ and $\beta$ only, as readily seen on Fig. \ref{sumfig}.
I obtain

\begin{eqnarray}
\frac{r_1}{y} &=& \sec{\alpha}
,
\label{r1case12}
\\
\frac{r_2}{y} &=& \sec{\beta}
,
\label{r2case12}
\\
\frac{r_{12}}{y} &=& 
\tan{\alpha} + \tan{\beta}
,
\end{eqnarray}

\noindent{}where $r_{12}$ indicates the distance between masses $M_1$ and $M_2$.

\subsection{Cases \#3-\#6}

\subsubsection{Masses}

Instead, for Cases \#3-\#6, I combined equations 36, 66-69, and 88 of \cite{erdczi2016}
to find

\begin{eqnarray}
\frac{M_1}{M} &=& \frac{
\tan{\beta} \left(\tan{\alpha} - \tan{\beta}\right)^2 \left(1 - 8 \cos^3{\beta}\right)
}
{
4\left[ \left(\sin{\alpha} - \cos{\alpha} \tan{\beta} \right)^3 -1\right]
}
\label{M1abCase3456}
\\
\frac{M_2}{M} &=& \frac{
\tan{\alpha} \left(\tan{\alpha} - \tan{\beta}\right)^2 \left(1 - 8 \cos^3{\alpha} \right)
}
{
4\left[ \left(\sin{\beta} - \cos{\beta} \tan{\alpha} \right)^3 + 1\right]
}
\label{M2abCase3456}
\end{eqnarray}

In Cases \#3-\#4, the angles $\alpha$ and $\beta$ are allowed to take on
values according to

\begin{eqnarray}
\alpha &=& 45^{\circ} + \kappa,
\\
\beta &=& \lambda \kappa
,
\end{eqnarray}

\noindent{such} that

\begin{equation}
0 \le \kappa \le 15^{\circ}
,
\label{kap34}
\end{equation}

\begin{equation}
0 \le \lambda \le 2
,
\label{lam34}
\end{equation}

\noindent{}whereas Cases \#5-\#6 instead require

\begin{eqnarray}
\alpha &=& 60^{\circ} + \kappa
,
\\
\beta &=& 60^{\circ} + \lambda \left(15^{\circ} - \kappa \right)
,
\end{eqnarray}

\noindent{such} that

\begin{equation}
0 \le \kappa \le 15^{\circ}
,
\label{kap56}
\end{equation}

\begin{equation}
-2 \le \lambda \le 0
.
\label{lam56}
\end{equation}

\subsubsection{Distances}

For all Cases \#3-\#6, the diagrams in Fig. \ref{sumfig} illustrate that

\begin{eqnarray}
\frac{r_1}{y} &=& \sec{\alpha}
,
\label{r1case3456}
\\
\frac{r_2}{y} &=& \sec{\beta}
,
\label{r2case3456}
\\
\frac{r_{12}}{y} &=& 
\tan{\alpha} - \tan{\beta}
,
\end{eqnarray}

\noindent{}where $r_{12}$ again indicates the distance between masses $M_1$ and $M_2$.

\begin{figure*}
\centerline{
\includegraphics[width=8cm]{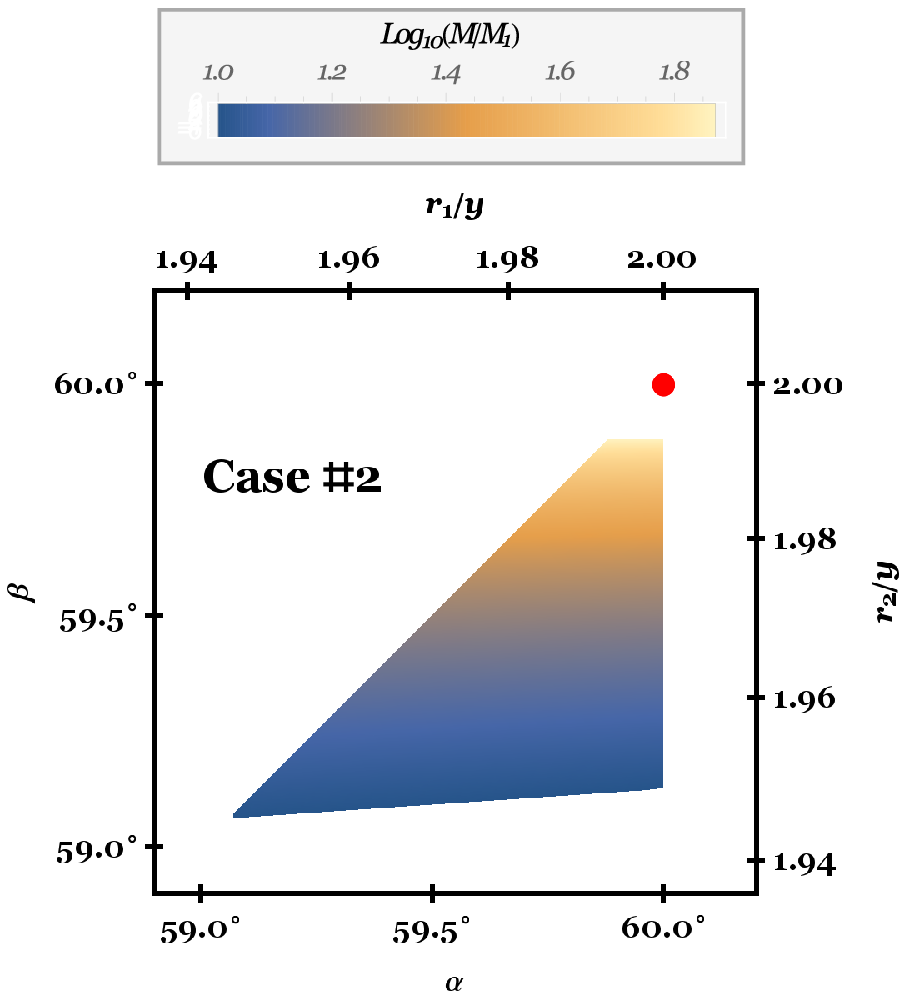}
\ \ \
\includegraphics[width=8cm]{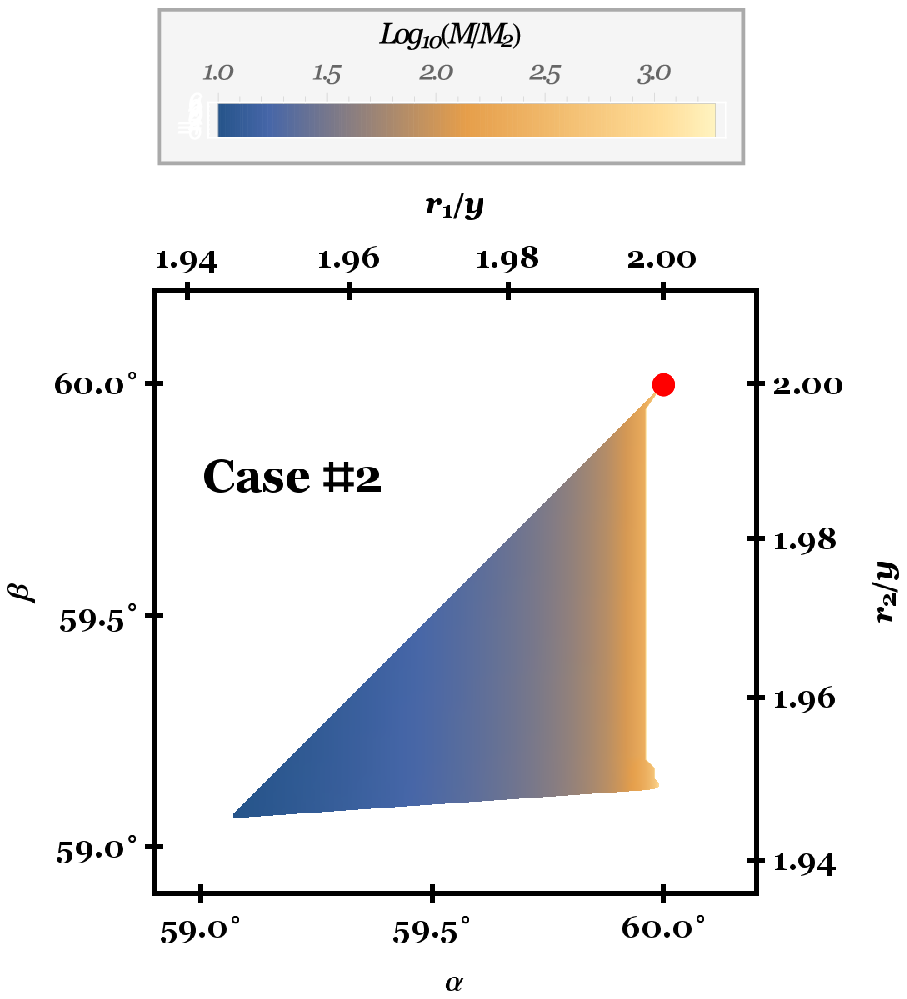}
}
\caption{
Same as Fig. \ref{case1mass}, but for Case \#2.
The allowed region for both $\alpha$ and $\beta$ is 
within a few degrees of $60^{\circ}$, and most
of the region is accessible only for large (super-Jupiter)
planets. Smaller objects such as asteroids would be confined
to the red dot.
}
\label{case2mass}
\end{figure*}

\subsection{Overall}

The last two sections explicitly illustrate the key insight of \cite{erdczi2016} that
the masses and distances between four planar bodies, two of which have equal masses, and
the other two defining an axis of symmetry, can be specified entirely by 
$\alpha$ and $\beta$. The conditions on $\alpha$ and $\beta$, however, require careful
attention, as they are interdependent. Further, the transcendental nature of equations 
(\ref{M1abCase12}-\ref{M2abCase12}) 
and (\ref{M1abCase3456}-\ref{M2abCase3456}) demonstrates the difficulty or impossibility
of obtaining explicit formulae for $\alpha$ and $\beta$ as a function of the masses.

The above results can be linked back to $\Lambda$ from equation (\ref{orig}) through
a transformation of equation 35 of \cite{erdczi2016}, giving

\begin{equation}
\Lambda = -\frac{M_1}{r_{1}^3}
          -\frac{M_2}{r_{2}^3}
          -\frac{1}{4} \left( \frac{M}{y^3} \right)
.
\label{lambnew}
\end{equation}

\noindent{}Equation (\ref{lambnew}) may be expressed as
the quantity $\Lambda y^3/M$ 
entirely in terms of $\alpha$ and $\beta$, helping to demonstrate
that the values of $y$ and $M$ effectively represent scaling
factors for all considered architectures.

\section{Link to real planetary systems}

Having established the computational method for this class of
central configurations, I now show how they may be applicable to
planetary systems.

\subsection{Mass cuts}

To do so, I perform mass cuts, and then implicitly solve the equations from Section 2
to determine regions of phase space in which solutions exist. In all cases, 
I assume that each star is at 
least ten times the mass of the substellar objects, and that the two stellar masses are within 
a factor of ten of each other.  These assumptions are realistic: the lowest-mass stars 
are a few tenths of a Solar mass, and the highest-mass planets are about an order of magnitude 
more massive than Jupiter ($\sim 0.01M_{\odot}$), roughly yielding a factor of ten in 
extreme mass ratio. All stars 
which are known to host planetary systems have masses between 
$0.13M_{\odot}-3.09M_{\odot}$ \footnote{http://exoplanets.org/}  
-- these stars include main sequence stars, giant branch stars, white dwarf stars and pulsars.

These assumptions also avoid some rare and ambiguous cases. For example, although a binary system consisting 
of a $100M_{\odot}$ star and $1M_{\odot}$ star might host a planetary system, none have
yet been observed, and would be extremely rare. Further, there is increasing evidence of a continuum of masses between
planets and stars through brown dwarfs such as L dwarfs, T dwarfs and Y dwarfs
\citep[e.g.][]{skretal2016}.  Although the definition of ``planet'' has become ambiguous, 
the boundary between planet mass and brown dwarf mass still lies in a restricted range from
about $0.010M_{\odot}$ to $0.015M_{\odot}$ \citep[e.g.][]{spietal2011}.

My assumptions provide no lower bounds on masses: they can reach zero (but must not be negative).
Depending on the level of accuracy sought, zero-mass objects could represent a variety of 
objects, such as dust grains, pebbles or asteroids.

\subsection{Results}

Implementing these mass assumptions immediately yields one result: no solutions exist
for Cases \#3 or \#5. This finding is indicated by red crosses in Fig. \ref{sumfig}.
Consequently, there exist no planar 4-body central configurations in binary-star 
planetary systems where the line of symmetry lies connects both stars and two
equal-mass objects are ``external'' to both stars.

For the remaining cases, indicated by green check marks in Fig. \ref{sumfig}, I created
three plots per case to summarize the results. Each plot is a density plot
as a function of the angles $\alpha$ and $\beta$ and distance ratios $r_1/y$ and $r_2/y$.
The angles and distance ratios are quantities which are interchangeable (see 
equations \ref{r1case12}-\ref{r2case12}
and \ref{r1case3456}-\ref{r2case3456}).  The density contours on the three
plots respectively display a mass ratio involving $M_1$, a mass ratio involving
$M_2$, and $r_{12}/y$. The red dots indicate extreme cases, where the mass ratios
become either zero or infinite. White patches between the red dots and the differently-coloured
regions indicate the gradient upon which this ratio takes a limiting value.
The colour scale for each plot is different in order to best showcase the
gradient across the allowed region.

\begin{figure}
\includegraphics[width=8cm]{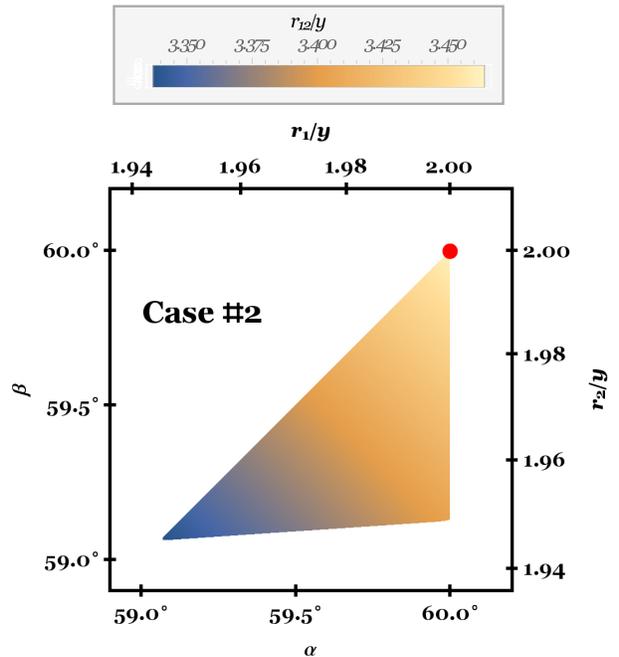}
\caption{
The distance $r_{12}$ between the substellar objects ($M_1$ and $M_2$),
for Case \#2.
This value, combined with the
allowed region for $\alpha$ and $\beta$, indicates
another near-rhombus-like configuration as in Case \#1.
}
\label{case2distance}
\end{figure}

\begin{figure*}
\centerline{
\includegraphics[width=8cm]{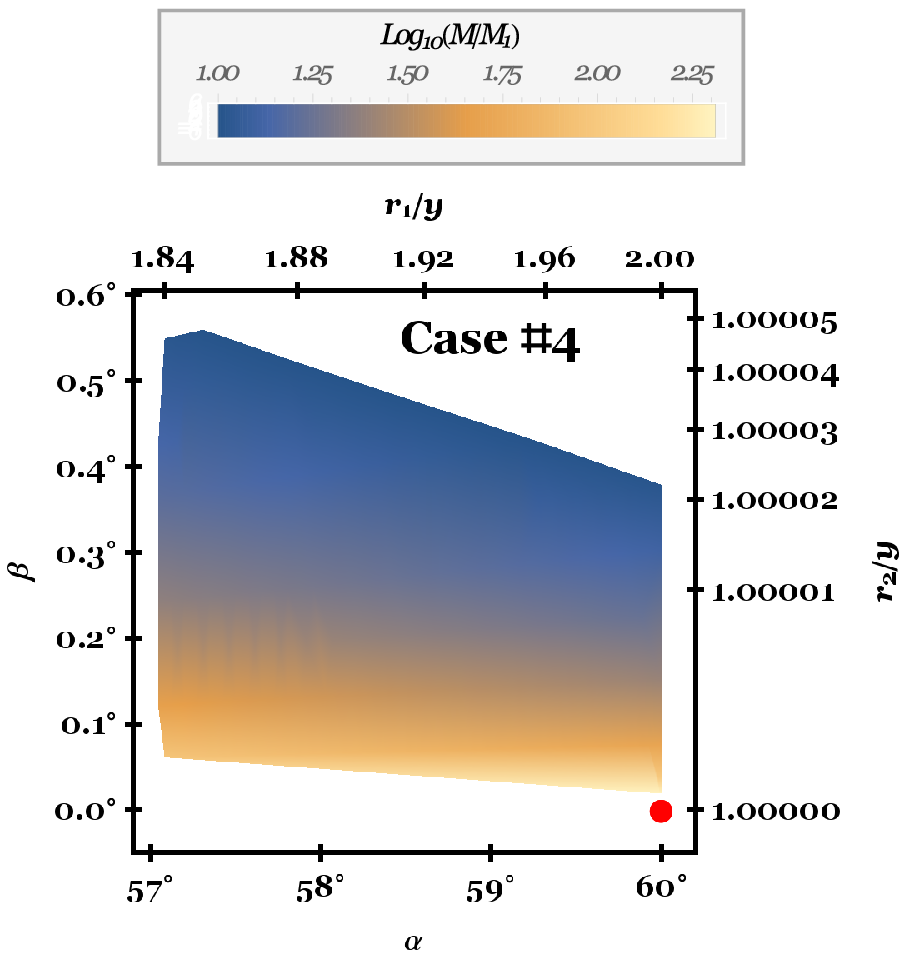}
\ \ \
\includegraphics[width=8cm]{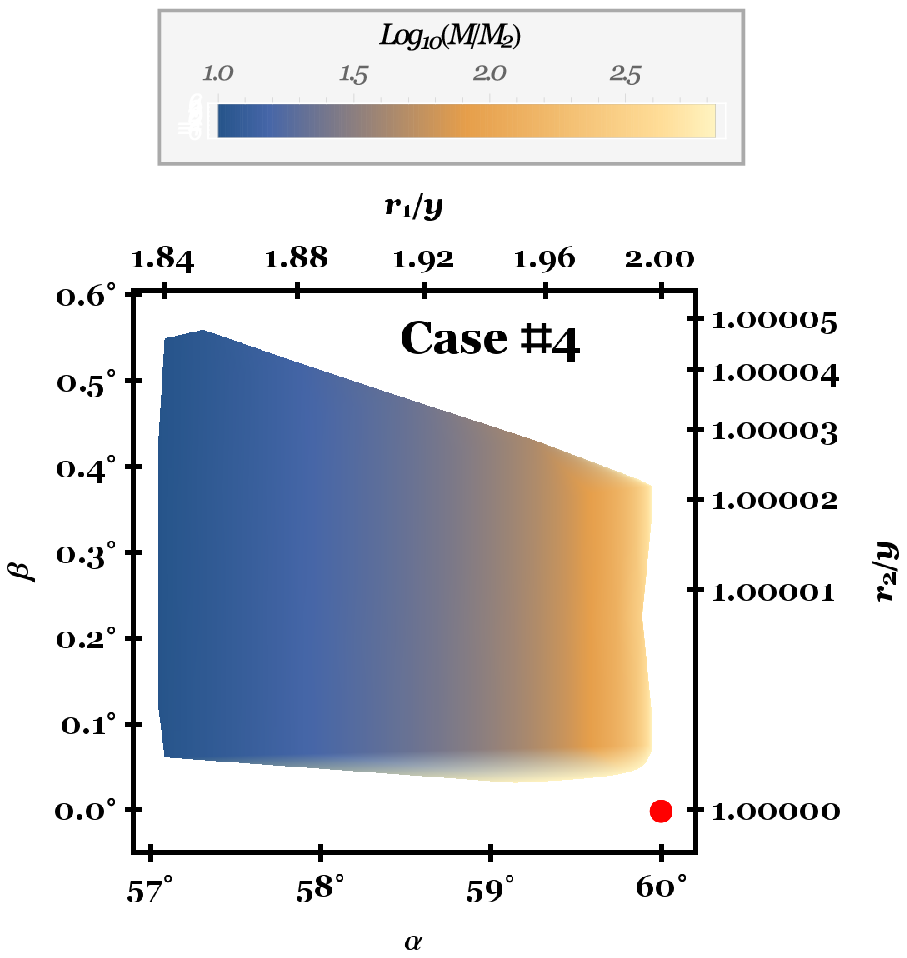}
}
\caption{
Same as Fig. \ref{case1mass}, but for Case \#4.
Here, because $\alpha \approx 60^{\circ}$ and 
$\beta \approx 0^{\circ}$, allowed
geometries include a near-co-linear configuration of both
stars plus one substellar body, with the other substellar body 
forming a near-equilateral triangle.
}
\label{case4mass}
\end{figure*}

\begin{figure}
\includegraphics[width=8cm]{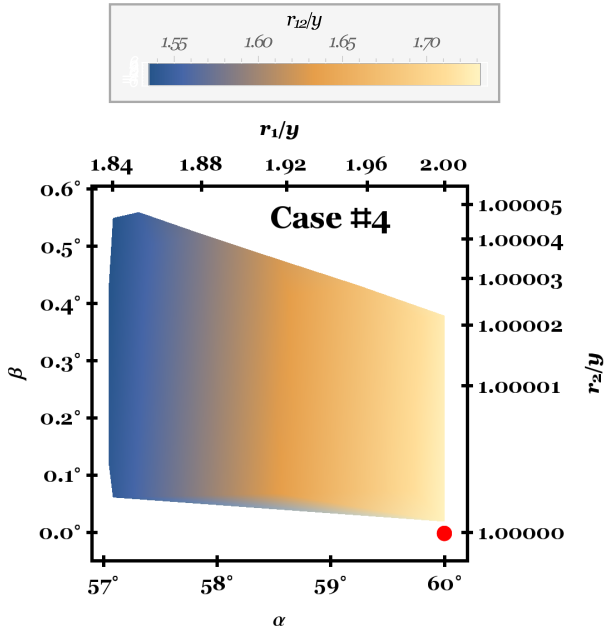}
\caption{
The distance $r_{12}$ between the substellar objects ($M_1$ and $M_2$),
for Case \#4.
}
\label{case4distance}
\end{figure}

\subsubsection{Case \#1 Results}

Case \#1 is the only unequal-mass star case that allows for central configurations.
I present the results in Figs. \ref{case1mass}-\ref{case1distance}.

One immediate striking observation is that the simple a priori restriction of 
$M_1/M$~$>$~$10$, $M_2/M$~$>$~$10$ and $10^{-1}$~$<$~$M_2/M_1$~$<$~$10$ reduces
the relevant $\alpha$ phase space range from $30^{\circ}$ 
(see equations \ref{alpha12} and \ref{kappa12}) to 
about $1.5^{\circ}$, and the $\beta$ phase space range
of $45^{\circ}$ 
(see equations \ref{beta12}, \ref{kappa12} and \ref{lambda12}) to 
about $1.3^{\circ}$. Another visually apparent
feature is that mass ratios of $M_1/M \sim 100$ hug the lower
boundary of the allowed region, whereas mass ratios of 
$M_2/M \sim 100$ instead approach the location of the red dot 
($\alpha = \beta = 30^{\circ}$), thereby showcasing the asymmetry 
produced from unequal stellar masses. 

Most of the allowed region
is then restricted to $10^2$~$>$~$M_1/M$~$>$~$10$ and $10^2$~$>$~$M_2/M$~$>$~$10$,
meaning that asteroid, pebbles or other small objects are confined
strongly to the location of the red dot.
The formal limit of both $M_1/M$ and $M_2/M$ at this value is infinity, 
meaning that arbitrarily small objects can exist in central configurations
there. 

As emphasized by Fig. \ref{case1distance}, the location of the red dot
indicates when an equilateral triangle configuration is formed between both
stars and either substellar object with mass $M$. This architecture then resembles a
rhombus with two internal angles of $120^{\circ}$ and the other two with
$60^{\circ}$. The stars would be at the vertices associated with the
$120^{\circ}$ angles.
 Effectively, this configuration is a doubled-up version of the
classic 3-body Lagrangian triangle, at least in terms of geometry.
In binary-star planetary systems, deviations from this rhombus
configuration extend only a few degrees in $\alpha$ and $\beta$, at most.

\subsubsection{Case \#2 Results}

I present the Case \#2 results in Figs. \ref{case2mass}-\ref{case2distance}.
In Case \#2, the stellar masses are equal and do not define the axis of symmetry.
Nevertheless, the inequality of the masses of the substellar objects still produce
a noticeable difference in the locations of the lightly-coloured regions 
between both panels of Fig. \ref{case2mass}.
Just as for Case \#1, the restriction of phase space due to the assumption of a 
planetary system is similarly stark: down to a range of about 
$0.9^{\circ}$ in both $\alpha$ and $\beta$. Only super-Jupiter mass planets
could exist in central configurations more than a few tenths of a degree
away from $60^{\circ}$ for either angle. 

The values of the mass ratios $M/M_1$ and $M/M_2$ both approach infinity as 
the red dot ($\alpha = \beta = 60^{\circ}$) is itself approached, indicating
that $M_1 \rightarrow 0$ and $M_2 \rightarrow 0$ is allowed. 
The red dot corresponds to a rhombus similar to that produced from the critical
red dot location in Case \#1, except rotated by $90^{\circ}$ and in the limit
that both stars have equal masses.

\subsubsection{Case \#4 Results}

Case \#4 is a fundamentally different geometry than Cases \#1 and \#2.
This difference is reflected in how the allowed phase
space in $\alpha$ and $\beta$ is restricted (Figs. \ref{case4mass}-\ref{case4distance}).
The value of $\alpha$ is restricted to $57^{\circ} - 60^{\circ}$ while $\beta$ is restricted to
$0.0^{\circ} - 0.57^{\circ}$. Effectively, these restrictions place $M_2$
nearly in-between both stars such that $M_1$ is in a near-equilateral
triangle configuration with the three near-co-linear bodies. At the critical point
($\alpha = 60^{\circ}$, $\beta = 0^{\circ}$), the values of $M_1$ and $M_2$ approach zero.
This combination of a co-linear configuration with an equilateral triangle
configuration represents a superposition of two three-body central configurations.

\subsubsection{Case \#6 Results}

\begin{figure*}
\centerline{
\includegraphics[width=8cm]{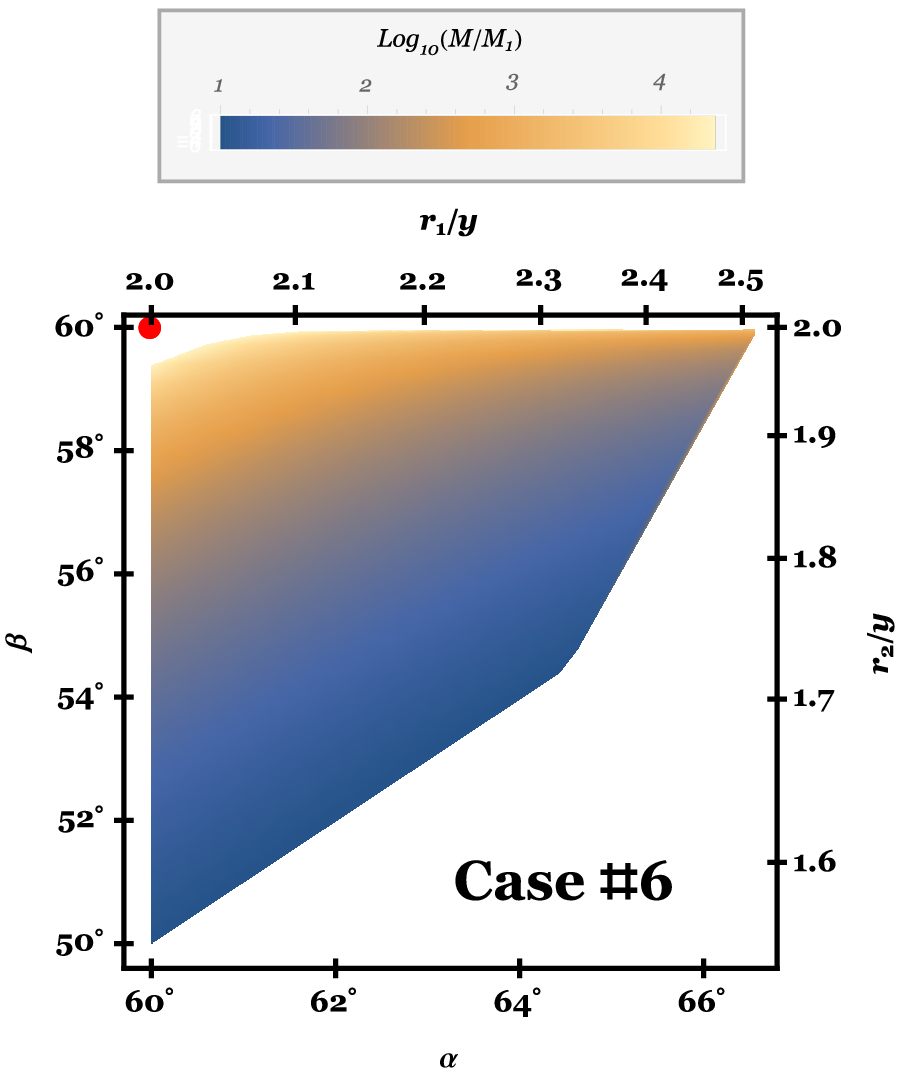}
\ \ \
\includegraphics[width=8cm]{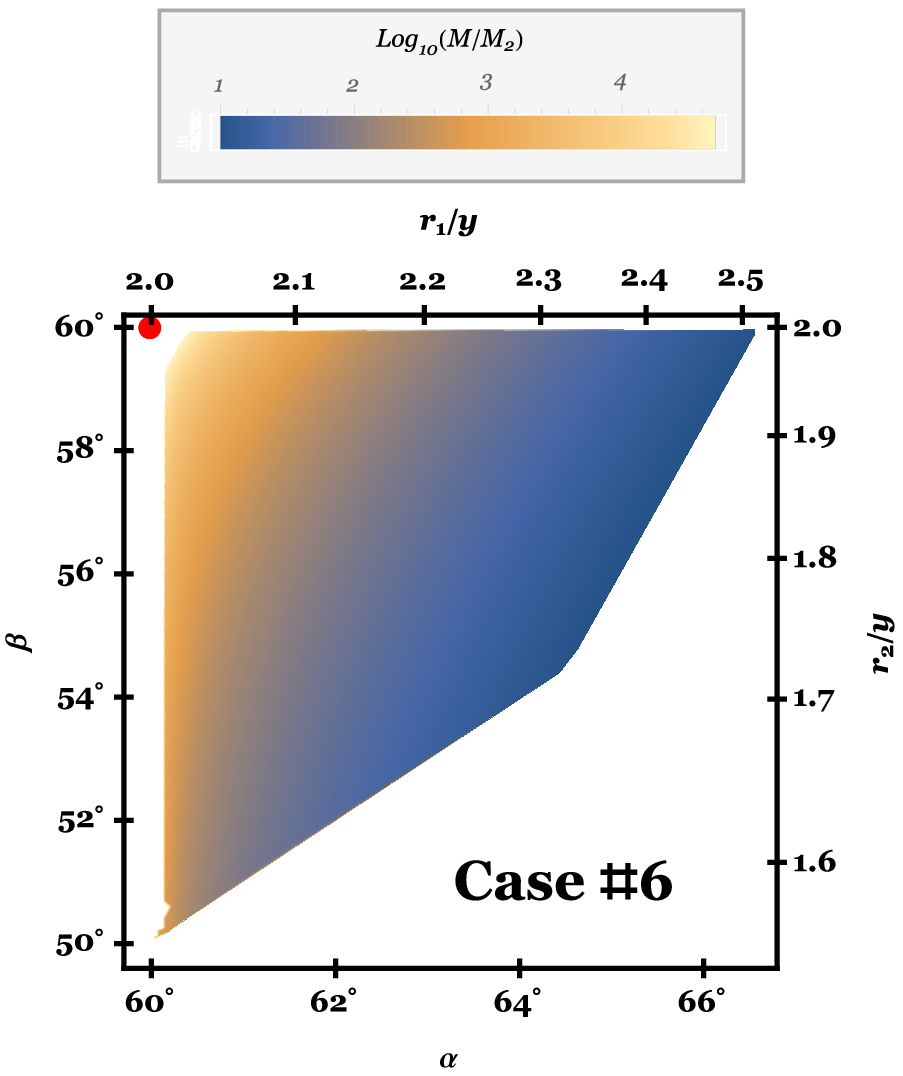}
}
\caption{
Same as Fig. \ref{case1mass}, but for Case \#6.
Here, $\alpha$ exceeds $60^{\circ}$ and 
$\beta$ is short of $60^{\circ}$. When the two angles
coincide, then an equilateral triangle would be formed
with one star at each vertex and both substellar masses
at the other.
}
\label{case6mass}
\end{figure*}

\begin{figure}
\includegraphics[width=8cm]{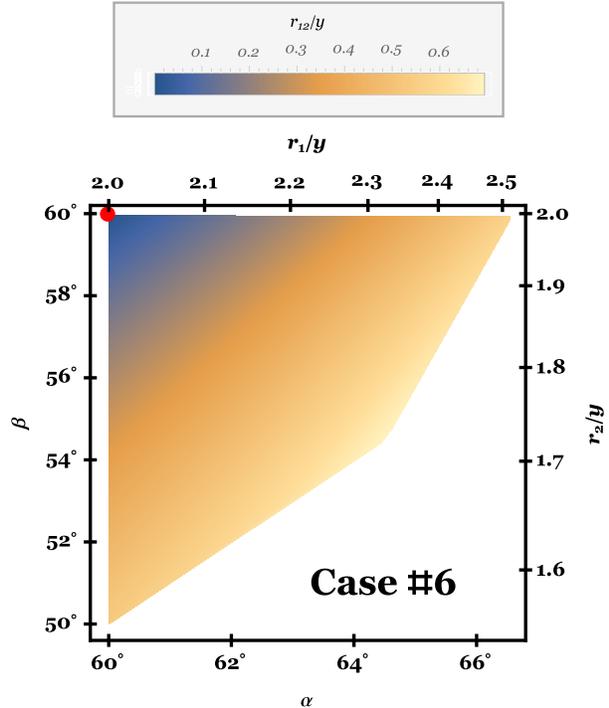}
\caption{
The distance $r_{12}$ between the substellar objects ($M_1$ and $M_2$),
for Case \#6.  Unlike for Cases \#1, \#2 and \#4, here $M_1$ and $M_2$
would coincide at the critical point.
}
\label{case6distance}
\end{figure}

Case \#6 is geometrically equivalent to Case \#4 except for the location
of the centre of mass. This difference creates a significant change in the
allowed $\alpha$-$\beta$ phase-space region (Figs. \ref{case6mass}-\ref{case6distance}),
with a range that extends to almost $10^{\circ}$ in $\beta$ and $7^{\circ}$ in $\alpha$.
At the critical point $\alpha = \beta = 60^{\circ}$, an equilateral triangle 
with one vertex that is ``doubled up'' with both $M_1$ and $M_2$ (which would coincide)
is formed. The kink which
appears in the diagonal on each plot indicates the point at which the 
$M = 10M_1$ and $M = 10M_2$ mass cuts were made.

\section{Discussion}

This study demonstrates that central configurations in binary-star
planetary systems may exist, and quantifies the extent of the existence
region. Whether objects residing in the immediate vicinity of these configurations are 
long-term stable and have plausible dynamical origins are different questions.

A proper stability analysis of planar four-body central configurations
with two equal masses and an axis of symmetry connecting the unequal masses
has not yet been carried out,
and could represent a significant undertaking. The results here might help
focus that effort by quantifying the variations in geometries and mass that
result from the most basic planetary system-based restrictions.

Regarding the likelihood of a planetary system forming or settling into
one of the central configurations discussed here, the chances probably
increase as the substellar masses decrease. All these central configurations include
objects which are nearly co-orbital or nearly co-linear. No planets
as massive as Jupiter have yet to be found in such architectures.
The largest known co-orbital objects in the solar system are Janus
($1 \times 10^{18}$ kg) and Epimetheus ($5 \times 10^{17}$ kg),
and in extrasolar systems is a minor planet with its disrupted 
fragments orbiting white dwarf 
WD 1145+017; the mass of this minor planet is likely to be about $10^{20}$~kg,
a tenth of the mass of Ceres \citep{guretal2016,rapetal2016}. Multiple
co-orbital masses above $10^{23}$~kg are roughly assumed to become unstable at 
the 20 per cent level \citep{veretal2016}.  Three or more 
co-linear planetary objects are not known to exist in stable configurations.

The fact that the majority of planets in the solar system host at least one
Trojan asteroid might indicate that capture into that central configuration
is common. If so, the extended Trojan-analogue rhombus configurations from
Cases \#1 and \#2 might also be easily populated by asteroids or smaller bodies
such as pebbles or dust or gas. In fact, remnant protoplanetary
disc features might reside in the locations indicated by the red dots on Figs.
\ref{case1mass}-\ref{case2distance}.

Finally, I emphasize that these configurations are fully scalable with
$y$ and $M$. The binary stars could have wide separations of $10^4$ au
or close separations of $10^{-2}$ au. Such close separations are unlikely
to host objects in central configurations because any bodies so close to those stars
would either be drawn in and destroyed due to tides, or blown away by winds.
For particularly wide separations, comparable to where stellar flybys
or planet-planet scattering might deposit planets or planetary debris into
an exosystem \citep{perkou2012,varetal2012}, capture into a central 
configuration is a distinct possibility. 

\section{Summary}

Central configurations are complete analytic solutions to the $N$-body problem and hence
allow one to determine the exact past and future evolution of a gravitational
system of point masses without the need for numerical integration. I have quantified
where in phase space do central configurations exist for coplanar four-body binary-star 
exoplanetary systems containing one axis of symmetry, two stars whose masses are 
within a factor of ten of each other, and two substellar bodies whose masses are
no greater than a tenth of either stellar mass.

I found that these basic mass restrictions 
-- 
which satisfy virtually every known planetary system
--
significantly restrict the geometries in which central configurations may occur
to within a few degrees of two architectures and ten degrees of a third. 
These architectures are all extensions of the well-known
observed Sun-Jupiter-Trojan configuration, and occur in the limit of
zero-mass substellar masses. The first architecture (Cases~\#1 and \#2) is a rhombus with both
stars at opposite ends whose associated internal angles are $120^{\circ}$. The second (Case~\#4) features
a co-linear configuration of two equal-mass stars plus one substellar body, with the second
substellar body forming an equilateral triangle. The third (Case~\#6) is an equilateral triangle
configuration with the two substellar masses at one vertex.

As the mass of substellar bodies
(which may be e.g. planets, asteroids, pebbles, dust) increases, the deviation from these
limiting architectures must increase in order to achieve the central configuration.
These mass restrictions also importantly exclude central configurations for the architectures
of Cases \#3 and \#5.  Binary stars need not be of equal mass nor classed as close nor 
wide in order to achieve central configurations.

\section*{Acknowledgements}

I thank the referee for their expert comments, particularly about Case \#6.  I have received funding from the European Research Council under the European Union's Seventh Framework Programme (BP/2007-2013)/ARC Grant Agreement n. 320964 (WDTracer).

\label{lastpage}
\end{document}